\documentclass[aps,floatfix,showpacs,twocolumn]{revtex4}
\usepackage{graphicx}

\begin{document}

\title{Galactic Potentials}
\author{Kayll Lake \cite{email}}
\affiliation{Department of Physics, Queen's University, Kingston,
Ontario, Canada, K7L 3N6 }
\date{\today}

\begin{abstract}
The information contained in galactic rotation curves is examined
under a minimal set of assumptions. If emission occurs from stable
circular geodesic orbits of a static spherically symmetric field,
with information propagated to us along null geodesics, observed
rotation curves determine galactic potentials without specific
reference to any metric theory of gravity. Given the potential,
the gravitational mass can be obtained by way of an anisotropy
function of this field. The gravitational mass and anisotropy
function can be solved for simultaneously in a Newtonian limit
without specifying any specific source. This procedure, based on a
minimal set of assumptions, puts very strong constraints on any
model of the ``dark matter".
\end{abstract}


\maketitle

There now seems to be wide spread belief that ``dark matter" is a
major constituent of our universe \cite{Bergstrom}. Whereas the
need for dark matter in some galactic halos has a long history
\cite{faber}, it is fair to say that at present we are quite far
away from a universally accepted solution to the dark matter
problem. Indeed, empirically motivated modifications of Newtonian
dynamics have been developed as an alternative to dark matter
\cite{mond}. Despite the profound implications of the problem, it
is essentially trivial to understand. Assuming the
non-relativistic Doppler effect and emission from stable circular
orbits in a Newtonian gravitational field, it follows that $V^2
\propto \frac{M(r)}{r}$ where $V$ is the orbital speed and $M(r)$
is the dynamical mass. But since $V^2$ levels off at large $r$ for
many galactic halos, $M$ must continue to grow like $r$. Since in
many cases the observed galactic components do not produce this
growth, what unseen material does?

The approach used here breaks the problem down into three steps:
(i) the determination of the galactic potential, (ii) the
construction of the effective gravitational mass from this
potential with the aide of an anisotropy function, and (iii) the
simultaneous solution of the effective gravitational mass and the
anisotropy function. Here we completely solve steps (i) and (ii)
for all metric type theories of gravity under a minimal set of
assumptions. Assuming only that emission occurs from stable
timelike circular geodesic orbits in a static spherically
symmetric metric with information propagated to us along null
geodesics, it is shown that the potential follows directly from
observed galactic rotation curves without any specific reference
to a theory of gravity. Further, without specifying any model of
the background, the introduction of an anisotropy function allows
the determination of the effective gravitational mass without
using Einstein's equations. Step (iii) is completed with the aide
of a Newtonian limit and the non-relativistic Doppler effect. This
last step makes it clear that the dynamical mass $M$ is not the
effective gravitational mass against which observed galactic
components should be compared to see if indeed any mass is
``missing" \cite{papers}.

\bigskip
First we construct stable circular timelike geodesic orbits in a
static spherically symmetric field \cite{relativity}.  In terms of
``curvature" coordinates the field takes the form \cite{metric}
\begin{equation}
ds^2=\frac{dr^2}{1-\frac{2m(r)}{r}}+r^2d\Omega^2-e^{2\Phi(r)}dt^2,
\label{standardform}
\end{equation}
where $d\Omega^2$ is the metric of a unit sphere. The use of
``curvature" coordinates plays no essential role in what follows
as it provides merely a basis for calculation. Throughout we refer
to the function $\Phi(r)$ as the ``potential" and $m(r)$ ($\neq
M(r)$) as the effective gravitational mass. A central point of
this analysis is the fact that this potential can be obtained
without knowledge of $m(r)$. It is immediately clear from
(\ref{standardform}) that all geodesic orbits are stably planar
(say $\theta=\pi/2$) and have two constants of motion, the
``energy" $\gamma=e^{2\Phi(r)}\dot{t}$ and ``angular momentum"
$l=r^2\dot{\phi}$ \cite{dot}. In the timelike case then
\begin{equation}
\dot{r}^2f(r)+\mathcal{V}(r)=\gamma^2, \label{effective}
\end{equation}
where
\begin{equation}
f(r)=\frac{e^{2\Phi(r)}}{1-\frac{2m(r)}{r}}\label{function}
\end{equation}
and
\begin{equation}
\mathcal{V}(r)=e^{2\Phi(r)}(1+\frac{l^2}{r^2}).\label{effectivepotential}
\end{equation}

\bigskip
Setting $\dot{r}=\ddot{r}=0,\; r>0$ it follows from the timelike
geodesic equations that \cite{henceforth}
\begin{equation}
\gamma=\frac{e^{\Phi}}{\sqrt{1-r\Phi^{'}}}\label{energy}
\end{equation}
and
\begin{equation}
l=\frac{r\sqrt{r\Phi^{'}}}{\sqrt{1-r\Phi^{'}}}\label{angularmomentum}
\end{equation}
where the time orientation has been chosen so that $\gamma>0$ and
all particles are assumed to rotate in the same sense with $\phi$
chosen so that $l>0$. Note that the existence of these circular
orbits requires
\begin{equation}
0<r\Phi^{'}<1.\label{orbits}
\end{equation}

 Next, we require that the timelike circular geodesics be
stable. Let $r_{0}$ be a circular orbit and consider
$r=r_{0}+\delta$ where $\delta<<r_{0}$. Taking expansions of
$\mathcal{V}(r)$ and $f(r)$ about $r=r_{0}$ it follows from
(\ref{effective}) that
\begin{equation}
\ddot{\delta}+\frac{\mathcal{V}^{''}(r_{0})}{2f(r_{0})}\delta=0\label{stable}
\end{equation}
so that $\mathcal{V}^{''}(r_{0})>0$ for stability. (The
requirement $\mathcal{V}^{'}(r_{0})=0$ merely reproduces
(\ref{angularmomentum})). From (\ref{effectivepotential}) then
\begin{equation}
3\Phi^{'}+r\Phi^{''}>2r(\Phi^{'})^{2}\label{stability}
\end{equation}
for stable circular orbits, a refinement of the Newtonian
condition $3\Phi^{'}+r\Phi^{''}>0$ for conservative central fields
\cite{stability}.

\bigskip
Under the assumption that information travels to us along null
geodesics, it follows, without further assumption,  that
\cite{sch}
\begin{equation}
1+z \equiv
\frac{\lambda_{o}}{\lambda_{e}}=\frac{(u_{\alpha}k^{\alpha})_{e}}{(u_{\alpha}k^{\alpha})_{o}}\label{redshift}
\end{equation}
where $\lambda$ is the wavelength, $e$ stands for the emitter, $o$
for the observer, $u_{\alpha}$ is the timelike four-tangent, and
$k^{\alpha}$ is tangent to the null geodesic (\textit{N})
connecting $e$ and $o$. The emitter is assumed to be on a stable
circular timelike geodesic in (\ref{standardform}). Along
\textit{N} define
\begin{equation}
b \equiv \frac{l_{N}}{\gamma_{N}}. \label{impactparameter}
\end{equation}
The constant $\gamma_{N}$ is positive by construction but $l_{N}$
is both positive and negative.  The observer is taken to be a
static observer at infinity
($u^{\alpha}_{o}=e^{-\Phi(\infty)}\delta^{\alpha}_{t}$).
Specifically, we assume that $\Phi(\infty)\rightarrow \mathcal{C}$
where $\mathcal{C}$ is a finite constant which we can set to zero
without loss in generality \cite{gauge}. (The ``fitting" problem
associated with the assumption that $\Phi(\infty)$ is finite is
discussed briefly below.) Then $b$ represents the impact parameter
at infinity. That is, $|b|$ gives the observed radial distance
either side of the observed center ($b=0$). The construction of a
mapping $b(r)$ (the mapping between the observer and coordinate
planes) is an important part of this analysis \cite{bending}.
Evaluation of (\ref{redshift}) now gives
\begin{equation}
1+z_{\epsilon}=\frac{1}{\sqrt{1-r\Phi^{'}}}(\frac{1}{e^{\Phi}}-\frac{\sqrt{r\Phi^{'}}\epsilon|b|}{r})
\label{redshift1}
\end{equation}
where $\epsilon=\pm1$. Rather than (\ref{redshift1}), we consider
shifts either side of the central value ($b=0$)  \cite{central}
\begin{equation}
1+z_{c}=\frac{1}{e^{\Phi}\sqrt{1-r\Phi^{'}}}. \label{redshift2}
\end{equation}
Defining
\begin{equation}
Z \equiv z_{+}-z_{c}=z_{c}-z_{-} \label{redshift3}
\end{equation}
we have
\begin{equation}
Z^2 =\frac {\Phi^{'}b^2}{r(1-r\Phi^{'})}. \label{redshift4}
\end{equation}

\bigskip
We now construct the mapping $b(r)$. At fixed $b$ (that is, at a
fixed offset from the observed center of the galaxy) choose the
maximum observed value of $Z$. From (\ref{redshift4}) it follows
that if $\frac {\Phi^{'}}{r(1-r\Phi^{'})}$ is monotone decreasing
with increasing $r$ then the maximum observed value of $Z$
corresponds to the minimum value of $r$ along $N$. This minimum
value follows from the null geodesic equation and is given by
\begin{equation}
b^2=\frac{r^2}{e^{2\Phi}}. \label{map}
\end{equation}
The monotone requirement gives us \cite{monotone}
\begin{equation}
\Phi^{'}>r\Phi^{''}+2r(\Phi^{'})^2. \label{monotone}
\end{equation}
With this restriction the mapping $b(r)$ is given by (\ref{map}).
\bigskip

Observations of galactic rotation curves are reported by way of
the ``optical convention"
\begin{equation}
v \equiv \frac{\lambda_{o}-\lambda_{e}}{\lambda_{e}}
\label{optical}
\end{equation}
so that
\begin{equation}
Z=v(b)-v(b=0) \label{equivalence}
\end{equation}
with Z given by (\ref{redshift4}) subject to the mapping
(\ref{map}). It is important to note that no ``velocity" has
entered the procedure \cite{velocity}. Note that $\Phi$ follows
directly from the observations, by way of a differential equation,
under a minimal set of assumptions: emitters on stable circular
orbits of the static field (\ref{standardform}) and emission along
null geodesics \cite{potential}. Whereas it has become customary
to decompose galactic potentials into various parts
\cite{persicsalucci}, such decompositions insert further
assumptions. We continue here with the full function $\Phi$ taken
as given directly from the observations assuming only that $v$ is
corrected for all systematic effects and reflects an intrinsic
property of the galaxy alone.

\bigskip
We now seek information on the function $m$. On the basis of the
Lovelock theorem \cite{lovelock} we know that properties of the
Einstein tensor are of central importance (even without invoking
Einstein's equations). Indeed, for spaces of the form
(\ref{standardform}) the entire structure of the space can be
specified by a single ``anisotropy" function
\begin{equation}
\mathcal{H} \equiv
G_{\theta}^{\theta}-G_{r}^{r}\label{hdef}\label{hhdef}
\end{equation}
where $G_{\alpha}^{\beta}$ is the Einstein tensor. It follows that
\cite{lake1}
\begin{equation}
m =\frac{\int \!b  {e^{\int \!a {dr}}}{dr}+\mathcal{C}}{{e^{\int
\!a {dr}}}} \label{massnew}
\end{equation}
where
\begin{equation}
a \equiv{\frac {2\,   {r}^{2}(\Phi^{''}+ (\Phi ^{'} )^{ 2})
 -3 r\,  \Phi^{'}
 \, -3}{r (  r \Phi^{'}
  +1 ) }},\label{a}
\end{equation}
and
\begin{equation}
b \equiv{\frac {r ( r (\Phi^{''}   +  (\Phi ^{'})
  ^{2}-\mathcal{H}) -\Phi^{'}  ) }{
 r \Phi^{'}   +1}},\label{b}
\end{equation}
with $\mathcal{C}$ a constant \cite{h}. Under the assumption of
spatial isotropy ($\mathcal{H}=0$) $m$ is determined, up to
quadrature, knowing $\Phi ^{'}$ and $\Phi ^{''}$. Moreover, any
function $m$ can be generated by a suitable choice for
$\mathcal{H}$. However, $m$ and $\mathcal{H}$ cannot be determined
simultaneously without further assumptions. Whereas it is a
straightforward matter to specify $\mathcal{H}$ via a specific
decomposition of the energy-momentum tensor (and many recent
papers do exactly this), such a procedure is not unique in the
sense that the same function $\mathcal{H}$ is derivable from
inequivalent decompositions. We proceed here in a different way by
invoking an assumption based on a Newtonian limit.

\bigskip
To place the forgoing analysis in Newtonian terms we introduce a
potential $\widetilde{\Phi}$ defined by \cite{landl}
\begin{equation}
\nabla^2 \widetilde{\Phi} \equiv -R_{t}^{t}
\label{newtonpotential}
\end{equation}
where $R_{t}^{t}$ is the time component of the Ricci tensor of
(\ref{standardform}) \cite{notnewton}. In the usual way
\cite{decompose} we find that $\widetilde{\Phi}$ satisfies
\begin{equation}
\widetilde{\Phi}^{'}=\frac{m}{r^2}-\frac{\Lambda
r}{3}+\frac{3}{r^2}\int_{0}^{r}(r \Phi^{'}-\frac{m}{r})dr +
\frac{1}{r^2}\int_{0}^{r}r^2\mathcal{H}dr ,\label{newtonian}
\end{equation}
where we have set $r>>2m$ and introduced the cosmological constant
$\Lambda$ \cite{newtonterms}. The dynamical mass is defined by $M
\equiv r^2 \widetilde{\Phi}^{'}$ and the balance of Newtonian
forces for circular motion gives
\begin{equation}
M=rV^2,\label{newtonianvelocity}
\end{equation}
where $V$ is the orbital speed. If the frequency shift is assumed
to be due to the non-relativistic Doppler effect then $V=v$ and
since $v$ is known, $\widetilde{\Phi}^{'}$ is known and we now
have two equations, (\ref{newtonian}) and (\ref{massnew}), from
which we determine a solution ($m$,$\mathcal{H}$). Note that no
decomposition of any energy-momentum tensor has been used.

\bigskip
Let us now define the energy density $8 \pi \rho \equiv -G_t^t$ so
that
\begin{equation}
m=\int_0^r 4 \pi r^2 \rho dr \label{standardmass}
\end{equation}
just as in Newtonian mechanics (though there $\rho$ stands for the
mass density). Equation (\ref{newtonian}) now provides the link
between the dynamical mass $M$ and the density $\rho$. The
traditional ``missing mass" problem derives from the fact that
since $v^2$ levels off at larger values of $r$ for many galactic
halos, assuming $V=v$ then according to (\ref{newtonianvelocity})
$M$ must continue to grow in that region like $r$. Assuming $M=m$
it then follows from (\ref{standardmass}) that there must be
unseen material if the inclusion of all observed contributions to
$\rho$ does not produce this continued growth. However, $M \neq
m$, and it is possible that the observed contributions to $\rho$
are compatible with $m$ while $M$ continues to grow like $r$
according to (\ref{newtonian}). In this sense there would be no
mass ``missing" at all. Mass should be considered ``missing" when
all observed contributions to $\rho$ are incompatible with $m$,
not $M$, and such an incompatibility could arise whether or not
$v^2$ levels off.

\bigskip
To finish we mention the fitting problem, but only briefly. This
involves the smooth junction of (\ref{standardform}) at a finite
value of $r$ (say $R$) onto an external field in which we can set
$\Phi(\infty)=0$. This boundary condition is required by the
definition of $b$ and by the fact that the rotation curves are
considered intrinsic (corrected for all other effects).
Geometrically this junction is examined by way of the
Darmois-Israel conditions \cite{musgrave}. The smooth junction of
metrics of the form (\ref{standardform}) only requires the
continuity of $m$ and $\Phi^{'}$, assuming the continuity of $r$,
$\theta$ and $\phi$, so that $G_r^r$, but not
$G_{\theta}^{\theta}$ nor $G_{t}^{t}$, is necessarily continuous
at $R$. In general relativity if the external field is taken as
vacuum then in a suitable gauge $e^{2\Phi}=1-2m(R)/r-\Lambda
r^2/3$ and so $\Phi(\infty)=0$ only for $\Lambda=0$.

\bigskip
In summary, assuming only that emission occurs from stable
timelike circular geodesic orbits in a static spherically
symmetric metric with information propagated to us along null
geodesics, it has been shown that galactic potentials follow
directly from the observed rotation curves via the relation
$Z^2=v^2$ where, subject to the mapping (\ref{map}), $Z^2$ is
given by (\ref{redshift4}). Neither the gravitational mass $m$ nor
any metric theory of gravity enters this determination of the
potential $\Phi$. Next, in terms of a single  anisotropy function
($\mathcal{H}$ given by (\ref{hdef})) $m$ follows directly from
$\Phi$. Both $m$ and $\mathcal{H}$ can be solved for, without
using any specific decomposition of the energy-momentum tensor, by
constructing a Newtonian limit and by now assuming that the
frequency shift is due to the non-relativistic Doppler effect.
This procedure naturally defines a dynamical mass $M \neq m$ with
respect to which the ``missing mass" problem is usually defined.
With $\Phi$, $\mathcal{H}$ and $m$ determined, the ``dark matter",
should it be required, is highly constrained, but not identified
since the same function $\mathcal{H}$ is derivable from
inequivalent decompositions of the energy-momentum tensor.

\bigskip
\begin{acknowledgments}
I thank Judith Irwin and Roberto Sussman for useful remarks. This
work was supported by a grant from the Natural Sciences and
Engineering Research Council of Canada. Portions of this work were
made possible by use of \textit{GRTensorII} \cite{grt}.
\end{acknowledgments}


\begin{thebibliography}{}
\bibitem{relativity}The procedures used here are ``relativistic" in
nature, not because of any claim that the ``dark matter" is a
priori necessarily a relativistic phenomenon, but rather because
the requisite tools are those in regular use for relativistic
phenomena. It should be noted that the first step (the use of
(\ref{standardform})) limits our considerations to metric type
theories of gravity.
\bibitem{metric}We use geometrical units throughout, a signature of +2 and the summation
convention.
\bibitem{gauge} Local linear translations in $t$ are physically irrelevant and
such transformations are equivalent to $\gamma \rightarrow
\gamma/D$ and $e^{2\Phi} \rightarrow e^{2\Phi}/D^2$ where $D$ is a
constant. As usual, $\Phi$ is determined only up to a constant.
Under such translations $1+z$ is invariant, as it must be
according to (\ref{redshift}).
\bibitem{bending}Another way to say this is that we are taking into
account any ``bending" of light.
\bibitem{central}It is instructive to start at the center ($b=0$) and examine
(\ref{redshift2}). Suppose that $1+z_{c}=1$ (which, it would
appear, is consistent with the observations) so that from
(\ref{redshift2})
\begin{equation}
e^{2\Phi}=1-cr^2 \label{centralredshift}
\end{equation}
where c is the constant of integration. Both (\ref{orbits}) and
(\ref{stability}) require $c<0$ and so the required potential is
of the anti-de Sitter form, though there is no suggestion that $c$
is universal. This already makes clear that there is a fitting
problem in the sense that if $\Phi(\infty)= 0$ then the
fundamental character of $\Phi(r)$ changes in the intervening
region.
\bibitem{velocity} To emphazise this point, there is no need
whatsoever to suggest at this point that the shift is due to the
non-relativistic Doppler effect. Relation (\ref{redshift})
includes all sources which, in any event, cannot be invariently
decomposed.
\bibitem{potential} This important point, often overlooked, is
stated explicitly by Nucamendi, Salgado and Sudarsky
\cite{papers}.
\bibitem{h}Equivalently, knowing $m$,
we can solve for $\mathcal{H}$ and obtain
\begin{equation}
-3 \frac{m}{r^3}+\frac{m^{'}}{r^2}-(1-\frac{2
m}{r})(\Phi^{''}+(\Phi^{'})^2)+\frac{\Phi^{'}}{r}(1-\frac{3
m}{r}+m^{'}).
\end{equation}
\bibitem{notnewton} Defining $8 \pi p_{_{R}} \equiv G_{r}^{r}$ we have
\begin{equation}
r \Phi^{'}=\frac{ 8 \pi p_{_{R}}r^3+2 m}{2(r-2
m)}.\label{definition}
\end{equation}
First consider $r>>2m$ (as in, for example, Mielke and Schunck
\cite{papers}). Then
\begin{equation}
r \Phi^{'} \simeq 4 \pi p_{_{R}}r^2 + \frac{m}{r}.
\label{intermediate}
\end{equation}
By inserting the exact form of $p_{_{R}}$ from (\ref{definition}),
(\ref{intermediate}) reduces to
\begin{equation}
 \Phi^{'} \simeq 0.\label{mands}
\end{equation}
A different approximation has been considered by Schunck
\cite{papers} (see Schunck's equation (1.9)). This approximation
is equivalent to
\begin{equation}
 e^{2\Phi} \simeq 1-\frac{2m}{r}.\label{schapprox}
\end{equation}
Neither of the approximations (\ref{mands}) nor (\ref{schapprox})
produce a workable Newtonian limit for our considerations here.
The construction (\ref{newtonpotential}) must be used.
\bibitem{decompose} Observe that
$R_{t}^{t}=G_{t}^{t}-\frac{1}{2}G-\Lambda$ from which
(\ref{newtonian}) follows without even a formal decomposition of
the energy-momentum tensor.
\end{thebibliography}
\end{document}